\documentclass[aps,prl,epsfig,showpacs,superscriptaddress,amsmath,amssymb,twocolumn]{revtex4-1}
\usepackage{multirow}
\usepackage{amssymb}
\usepackage{amsmath}
\usepackage{latexsym}
\usepackage{graphicx}
\usepackage{dcolumn}
\usepackage{bm}
\usepackage{ulem}
\usepackage{color}
\usepackage{soul}
\graphicspath{ {./figs/} }

\newcommand{\be}{\begin{equation}}
\newcommand{\ee}{\end{equation}}
\newcommand{\ba}{\begin{eqnarray}}
\newcommand{\ea}{\end{eqnarray}}
\newcommand{\baa}{\begin{eqnarray*}}
\newcommand{\eaa}{\end{eqnarray*}}
\newcommand{\dg}{^{\dagger}}
\newcommand{\mI}{{\mathbf 1}}
\newcommand{\bk}{{\bf k}}

\newcommand \renp{\renewcommand{\multirowsetup}{\centering}\multirow}
\newcommand \zx{\hbox{zx}}
\newcommand \zy{\hbox{zy}}

\begin{document}

\title{TAO pairing: a fully gapped pairing scenario for the Iron-Based Superconductors}
\author{T.~Tzen Ong}
\affiliation{Center for Materials Theory, Department of Physics \& Astronomy, Rutgers University, Piscataway NJ 08854, USA}
\author{Piers Coleman}
\affiliation{Center for Materials Theory, Department of Physics \& Astronomy, Rutgers University, Piscataway NJ 08854, USA}
\affiliation{Department of Physics, Royal Holloway, University of London, Egham, Surrey TW20 0EX, UK}
\date{\today}

\begin{abstract}
Motivated by the fully gapped superconductivity in iron-based
superconductors with uncompensated electron pockets, we propose a spin
singlet, but orbital triplet analogue of the superfluid phase of
${}^3$He-B.
We show that orbital triplets with a nominal d-wave
symmetry at the iron sites can transform as s-wave pairs
under rotations about the selenium sites.  Linear
combinations of such d$_{xy}$ and d$_{x^{2}-y^{2}}$ triplets form a
fully gapped, topological superconductor.
Raman-active excitations are predicted to develop
below the superconducting transition temperature.
\end{abstract}
\maketitle

%\section{Introduction}

% 1. Gen intro
% 2. Big questions about the Coulomb interaction.
% 3. He-3B as motivation for a new gap model.
% 4. Introduction to the new quantum numbers, table and pictures.

The discovery of superconductivity with $T_c = 26$K in LaFeAsO by
Hosono {\it et. al.}\cite{HosonoJACS08} has opened a new field of
iron (Fe)-based  multi-band high temperature superconductors
(SCs)\cite{MazinRepProgPhys11,PaglioneNPhys10}.
STM, ARPES and bulk experiments show that a majority of these systems these systems are fully gapped\cite{EisakiTerasakiJPSJ09,EisakiYashimaJPSJ09,MatsudaPRL2009,MolerJPSJ09,KAMPRL2011,MazinRepProgPhys11,HanaguriScience10,HoffmanRPP11,AllanScience12,HongEPL12}
while optical measurements also show a
significant Coulomb interaction\cite{BasovNPhys09}.
Most electronically mediated superconductors,
avoid the on-site Coulomb interactions by building nodes
into the pair wave, so that the on-site pairing
$\langle \psi_{\uparrow}\psi_{\downarrow} \rangle_{\rm Fe} = \int_{\bk }\frac{\Delta_{\bk }}{2E_{\bk }} = 0$.
In the organic, heavy fermion, cuprate and strontium ruthenate
superconductors, this ``Coulomb orthogonalization'' is guaranteed by a d- or p-wave
symmetry of the order parameter\cite{Jerome2008,BatailPRL00,PfleidererRMP09,Tsuei2000,Mack03}.

Superconductivity in the Fe-based systems is widely believed to derive
from a gap function with $s^\pm$ symmetry
\cite{mazinschmalian2009,MazinRepProgPhys11}, taking the form
$\Delta (\bk)= \Delta_0 + 2 \Delta_1 \cos k_x \cos k_y$,
containing a node between the electron and hole pockets.
This is supported by c-axis Josephson
tunneling experiments\cite{TakeuchiPRL09}, QPI measurements on Fe(Se,Te)
\cite{HanaguriScience10}, and the absence of the Wohlleben effect
\cite{MolerJPSJ09}.  In this scenario, a phase
cancellation between the electron and hole pockets eliminates on-site
pairing, minimizing the Coulomb interaction. However, when the hole
pockets are absent, such as in A$_x$Fe$_2$Se$_2$ \cite{FengNatMat11}
and single layer FeSe \cite{ZhouNatComm12}, the node in the $s^{\pm}$
order parameter is expected to intersect the the electron pockets to
achieve Coulomb orthogonalization. These nodes have not been
observed. One possibility is that the symmetry of the order
parameter changes in the electron pocket materials, for example by the
development of d-wave pairing, with nodes between the pockets
\cite{ScalapinoHirschfeldPRB11}; yet this scenario appears to be
inconsistent with the observation of fully gapped electron pockets
around the $Z$-point in Tl$_{0.63}$K$_{0.37}$Fe$_{1.78}$Se$_2$
\cite{HongEPL12}.

These observations motivate the search for an alternative description
of the pairing symmetry to account for the Coulomb
orthogonalization in single
electron-pocket materials without a change in condensate symmetry.
Here, we are inspired by
superfluid $^{3}$He-B, where the Fermi surface is fully gapped, yet the
pair wavefunction contains hidden nodes.
In two-dimensional $^{3}$He-B, the gap function takes the form
$\langle c_{\bk \sigma }c_{-\bk \sigma '}\rangle =
(k_{x}\sigma_{x}+k_{y}\sigma_{y})i\sigma_{2}$;
although the odd-parity $k_{x}$ and $k_{y}$ triplet
components contain nodes, their anti-commuting spin structure
causes them to add in quadrature to create a fully gapped condensate where
$\Delta (\bk )\propto [k_{x}^{2}+k_{y}^{2}]^{1/2}$.
While spin-triplets are ruled out in
iron-based superconductors by Knight-shift measurements
\cite{EisakiTerasakiJPSJ09,EisakiYashimaJPSJ09},
here we show that the multi-orbital nature of the iron-based superconductors
opens up an analogous class of non-trivial orbital triplet states.

\begin{figure}[bht!]
\begin{center}
\includegraphics[trim=0mm 0mm 0mm 1mm, clip, width=\columnwidth]{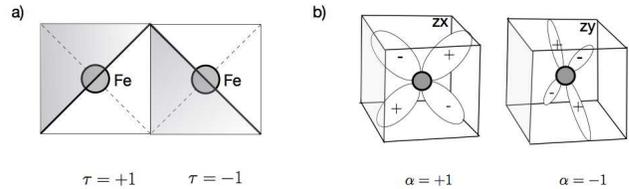}
\end{center}
\caption{\label{Staggeredtetrahedra} (a) Tetrahedral index
$\tau =\pm 1$ denotes Fe in right or left pointing tetrahedra,
(b) orbital index $\alpha =\pm 1$ denotes zx and zy orbitals.}
\end{figure}

The staggered tetrahedral structure of the iron based
superconductors results in an enlarged unit cell containing two
rotated Fe tetrahedra.
Electrons on iron sites in ``right'' or ``left'' pointing
tetrahedra (Fig.~\ref{Staggeredtetrahedra}a.)
are labelled by a ``tetrahedral'' band index $\tau= \pm
1$ and the Pauli matrices linking these states will  be denoted by
$\vec{\tau }= (\tau_{1},\tau_2,\tau_3)$.
Likewise, we associate an orbital index $\alpha =\pm 1$ with the degenerate
{\sl zx} ($\alpha=+1$) and {\sl zy} ($\alpha =-1$) iron d-orbitals, denoting
the Pauli matrices that link these states by $\vec{\alpha }=
(\alpha_{1},\alpha_{2},\alpha_{3})$ (Fig. \ref{Staggeredtetrahedra}b).
%\begin{equation}\label{}
%\figx{3truein}{figs/fig1_tetra.eps} \nonumber
%\end{equation}
\vskip -0.2in
\noindent The electron operators now carry three discrete quantum numbers,
so we replace the conventional electron operator
$c_{\bk \sigma}\rightarrow c_{\bk \nu}$, where  $\nu\equiv
(\tau ,\alpha ,\sigma )$ denotes the triplet of site, orbital and spin indices.
Since we are treating orbital triplets, we adopt a generalized
Balian-Werthamer spinor to describe the electron fields
\begin{equation}
\Psi_{\bk} \equiv {\psi_{\bk\nu} \choose -i (\lambda_{2})_{\nu\nu'}
 \psi^{\dagger}_{-\bk\nu'}}
\label{eq:BW spinor}
\end{equation}
where $\lambda_{2} = \tau_{2} \alpha_{2} \sigma_{2} $ flips the
tetrahedral, orbital and spin quantum numbers so that so that the electron and hole fields transform the same way under isospin rotations.
In this notation an extended $s^{\pm}$ pair,  written
$\Delta^{s \pm}(\bk) = cos(k_x) cos(k_y) \tau_{2} \alpha_{2}
\mI_{\sigma}$,  is an orbital and site triplet,
the orbital analog of equal spin triplet pairing,
for once multiplied by $\lambda_{2}$, it is
diagonal in site and orbital indices.
We now consider a more general class of
orbital triplet pairing.

The 2D Fe As(Se) layer exhibits C$_{4v}$ symmetry about the
As(Se)atoms, and D$_{2d}$ symmetry around the Fe atoms,
{with an inversion center on the Fe-Fe bond}.
We choose to use the C$_{4v}$ symmetry as it naturally accounts for the two Fe atoms per unit cell.
There is no $90^{\circ}$ rotation symmetry about an
iron site - instead, the unit cell is invariant under a combined
$90^{\circ }$ rotation and a translation between the two iron sites.
This means that the site and orbital indices
transform non-trivially under symmetry operations of the C$_{4v}$
group. For example, under a $90^{\circ }$ rotation,
$\vert \zx\rangle
\rightarrow \vert \zy\rangle $,
$\vert \zy\rangle
\rightarrow -\vert \zx\rangle $, so that
the orbital matrix
$\alpha_{1}\equiv  \vert \zx\rangle \langle  \zy\vert  + \vert \zy
\rangle  \langle \zx\vert $ changes sign, $\alpha_{1}\rightarrow - \alpha_{1}$.
This means that the global symmetry of the superconducting (SC) state need
not correspond to the momentum-space  symmetry of the pairs.
For example, the orbital triplet $(k_{x}^{2}-k_{y}^{2})
\alpha_{1}$ involving the product of
a gap of nominal d-wave symmetry
and an orbital field $\alpha_{1}$
forms an s-wave pair, because both $k_{x}^{2}-k_{y}^{2}$ and
$\alpha_{1}$ change sign under a $90^{\circ }$
rotations \cite{FCZhangPRB2008,WangEPL2009}.

We shall argue that the strong Coulomb forces at
the iron sites favor states with local $d$-wave symmetry,
seeking gap functions which contain
terms with both local $d_{xy}$ and
$d_{x^{2}-y^{2}}$ spatial symmetries coupled to different site and
orbital channels. Previous analyses have considered
pairing states that involved a single orbital
channel\cite{FCZhangPRB2008, WangEPL2009, Daiarxiv0805.0736}. Here
we seek pairing states which involve an entangled combination of
two different orbital channels.
The orbital triplet pair wave-function takes the  form
\begin{equation}
\langle c_{\bk \nu} c_{-\bk \nu' }
\rangle = (\Delta_{xy}(\bk) {\Gamma}^1_{\nu\nu''} + \Delta_{x^{2}-y^{2}} (\bk )
\Gamma^2_{\nu\nu''} )(i \lambda_{2})_{\nu'' \nu'},
\label{eq:SC order parameter}
\end{equation}
where $\Delta _{xy} (\bk )= \Delta_{1}\sin k_{x}\sin k_{y}$ and
$\Delta _{x^{2}-y^{2}} (\bk )= \Delta_{2}(\cos k_{x}-\cos k_{y})$
are the $xy$ and $x^{2}-y^{2}$ pair amplitudes and the
$\Gamma^{1,2}\in [\tau_{a}\otimes \alpha_{b}]$ are
tensor products of the site and orbital Pauli matrices.
These spin singlets are  odd parity under spin
exchange, but even under momentum
inversion, site ($\tau $) and orbital ($\alpha $) exchange, thus preserving
the antisymmetry of the wavefunction.

To preserve lattice symmetry, $\Gamma_{1}$ and $\Gamma_{2}$ must belong
to the {\sl same} irreducible representation of the C$_{4v}$ group and
in addition, both must have the same parity under time-reversal and inversion.
Most importantly, the two d-wave
pairing components of the Hamiltonian must anti-commute with each other,
to guarantee that their gaps add in quadrature.
We call this ``tao pairing'' (tetrahedral and orbital: $\tau \alpha \sigma$).
One of its features is that as in $^{3}$He-B,
the d-vector $\vec{d} (\bk) = (\Delta_{xy} (\bk ),\Delta_{x^2-y^2} (\bk ))/|\Delta (\bk)|$
has {\sl topological} character.
In particular, the off-diagonal
components of the pairing in orbital space
entangles the orbital components
of the pair wavefunction so that the winding number
\begin{equation}\label{eq:topological number}
\oint_{\Gamma} \left( \vec{ d}\ \dg  (\bk )\times \nabla_{\bk }\vec{d} (\bk )\right)
\cdot d\bk = 2 \pi \nu,
\end{equation}
evaluated around a path that encloses the $\Gamma$ point,
is a topological invariant.
The ``chirality'' $\nu =- {\rm sgn} (\Delta_{1}\Delta_{2})$
of the order parameter is selected by the
underlying Hamiltonian without breaking any symmetry.
\begin{center}
\begin{table}[bt]
\begin{tabular}{|c|c|c|c|} \hline
Irreducible & Gap Fn. &
Local  & Orbital ($\alpha $)\\
Rep. &&Symmetry&Symmetry
\\ \hline
$A_{1}$ & $c_{x}c_{y} \ \tau_{2}\alpha_{2}$ &
 $s^{\pm}$& $11+22$
\\ \hline
$B_{1}$ & $i\Delta_{x^2-y^2}\  \tau_{3}\alpha_{2}$ & $d_{x^2-y^2}$&
$11+22$
\\ \hline
$B_{2}$ &
$\Delta_{xy}\ \tau_{2} \alpha_{2}$ &
$d_{xy}$ & $11+22$ \\ \hline \hline
\renp{2}{10mm}{$A_{1}$}
 &
$\Delta_{x^2-y^2} \tau_3 \alpha_1 $ &  $d_{x^2-y^2}$ & $11-22$ \\
\cline{2-4}
 &
$i\Delta_{xy} \tau_{2} \alpha_3 $ &  $d_{xy}$ & $12+21$ \\
\hline \hline
\renp{2}{10mm}{$A_{2}$} &
$\Delta_{x^2-y^2} \tau_3 \alpha_3 $ &
 $d_{x^2-y^2}$ &$12+21$ \\ \cline{2-4}
&$i\Delta_{xy} \tau_{2} \alpha_1 $ & $d_{xy}$ & $11-22$ \\ \hline \hline
$A^{*}_{1}$ & $i\Delta_{xyz} \tau_1 \alpha_{2} $ &
3D staggered $d_{xy}$
 & $11+22$ \\
&&   (``$\eta $'' )
&\\
\hline
$B^{*}_{1}$ & $\Delta_{xyz} \tau_1 \alpha_1$ &
3D staggered $d_{xy}$ &$11-22$ \\ \hline
\end{tabular}
\caption{Symmetry classification of the singlet superconducting order
parameters in the Fe-based systems. The non-trivial s-wave
representations $A_{1}$ and
$A_{2}$ contain two independent entries
involving a product  of a d-wave gap function
with  an orbital field $\alpha_{1,3}$ which also changes sign under
$90^{\circ }$ rotations. The  $A_{1}^*$ and $B^{*}_{1}$ representations
involve  staggered $d_{xy}$ order.
The $\tau_{1}$ operator which implements the staggered pairing is
odd-parity under inversion,
so that the ``f-wave'' combination
$\Delta_{xyz}\tau_{1}= (\Delta_{xy} \tau_{1}) \times  \sin k_z$ has
{\sl even} parity.
Since $\alpha_{2}$ and $\tau_{2}$ are odd
under time-reversal, the time-reversal invariant combinations
$i\alpha_{2}$ and $i\tau_{2}$ are used in the pair operator.
}
\label{table:SCsymmetry}
\end{table}
\end{center}
\vskip -0.2truein

Table.~\ref{table:SCsymmetry} classifies the
even-parity order parameters, with nearest or next-nearest
neighbor pairing in the Fe-based SCs according
to the irreducible representations of the C$_{4v}$ group.
Within the s-wave $A_{1}$ and $A_{2}$ symmetry classes there are
two one dimensional representations with
local $d_{xy}$ and local $d_{x^{2}-y^{2}}$ symmetry, in which the
product of an orbital field $\alpha_{1,3}$ with a d-wave gap function
creates an s-wave pair.
Their identical symmetries under C$_{4v}$ means
that hopping terms in the Hamiltonian
inter-convert the two types of pair, creating
an internal Josephson coupling between them.

In this paper, we focus on the
$A_{1}$ and $A_{2}$ states, which
are  invariant under inversion symmetry of a single layer.
The $A^{*}_{1}$ and $B^{*}_{1}$ representations,
which involve staggered $d_{xy}$ pairing
are odd-parity within a single plane and
can only form a  tao state in
bulk 3D systems by staggering the $d_{xy}$  pairing between
layers; in three dimensions, the $A^*_{1}$ and $B^*_{1}$ states acquire an additional  $\sin k_z$ dependence,
{restoring their even parity} giving rise to the $\Delta _{xyz}$ =
$(\mathrm{staggered} \, \Delta _{xy}) \times \sin k_z$ form-factor.
We note that the $A^{*}_{1}$ state corresponds to the $\eta$-pairing
proposed in \cite{HuPRX12}.

Using this scheme, we are able
to identify {\bf two} fully gapped 2D tao states,  with
$A_{1}$ and $A_{2}$ symmetry (see Table.~\ref{table:Fully gapped SC states}).
In multi-layered iron-based systems, the possibility of introducing a
dependence of the gap on the momentum $k_{z}$ perpendicular to the layers,
allows us to mix the ``starred"representations, which have a $sin k_{z}$
dependence, with the ``unstarred" representations.
\begin{center}
\begin{table}[tb]
\begin{tabular}{|c|c|c|} \hline
Gap Function & Symmetry
& Description \\
\hline
$
\Delta_{xy} \tau_{2} \alpha_3 \gamma_{2}+
\Delta_{x^2-y^2} \tau_3 \alpha_1 \gamma_{1}
$
& $A_{1}$  & Uniform\\
\hline
$
\Delta_{xy} \tau_{2} \alpha_1 \gamma_{2}
+
\Delta_{x^2-y^2} \tau_3 \alpha_3 \gamma_1
$ & $A_{2}
$  &  Uniform\\
\hline \hline
$\Delta_{xyz} \tau_1 \alpha_{2} \gamma_{2} + \Delta_{x^2-y^2} \tau_3 \alpha_1 \gamma_{1}$ & $A^{*}_{1}$ &  Staggered $d_{xyz}$. \\
\hline
$\Delta_{xyz} \tau_1 \alpha_1 \gamma_1 + \Delta_{x^2-y^2} \tau_3 \alpha_{2} \gamma_{2}$ & $B^{*}_{1}$ &  Staggered $d_{xyz}$. \\
\hline
\end{tabular}
\caption{Family of orbital triplet tao superconducting states. The asterix denotes
tao-states which break the inversion symmetry of a single FeAs layer.
The Nambu matrices acting in particle-hole space are denoted by
$\vec{\gamma}= (\gamma_{1},\gamma_{2},\gamma_{3})$.
}
\label{table:Fully gapped SC states}
\end{table}
\end{center}

To illustrate these results, we now carry out a BCS treatment of
the $A_{1}$ tao state \ref{table:Fully gapped SC states}), with
Hamiltonian
\begin{equation}
H = \sum_{\bk }\Psi\dg_{\bk }h (\bk )\Psi_{\bk }
- \frac{g_{1}}{N_{s}}
\sum_{\bk ,\bk' }
b\dg_{\bk 1 }b_{\bk'1 }
- \frac{g_{2}}{N_{s}}
\sum_{\bk ,\bk' }
b\dg_{\bk 2 }b_{\bk' 2},
\label{eq:BCS Model}
\end{equation}
where  $N_{s}$ is the number of sites in the lattice and
\begin{eqnarray}
b\dg _{\bk 1} &=&
c\dg_{\bk \nu}
\biggl[
\biggl(
d_{xy} (\bk ) i\tau_2\alpha_3
 \biggr)
 i\underline{\lambda}_{2}\biggr]_{\nu\nu'}c\dg_{-\bk \nu'},\cr
b\dg _{\bk 2} &=&
c\dg_{\bk \nu}
\biggl[
\biggl(
d_{x^{2}-y^{2}
} (\bk )\tau_3 \alpha_1
 \biggr)
 i\underline{\lambda}_{2}\biggr]_{\nu\nu'}c\dg_{-\bk \nu'}
 \label{eq:A1g order parameters}
\end{eqnarray}
create the two components of a tao-pair with $A_{1}$ symmetry,
where $d_{xy} (\bk)$ and $d_{x^{2}-y^{2}} (\bk)$ are the $d$-wave form
factors.
We note that a microscopic model is most likely
to involve orbital and spin superexchange processes, giving rise
to Kugel-Khomskii type $J_{1}-J_{2}$ interactions \cite{KugelSPU82}.
For simplicity, we have omitted interaction terms that mix the
the two pair types, since as we shall see, this mixing is already
provided by the hopping.

The kinetic part of the Hamiltonian
\begin{equation}
h (\bk )=  \epsilon_{0}(\bk)  \gamma_{3}  + \epsilon_{1}(\bk)
\tau_1 + \epsilon_{3}(\bk) \alpha_1 + \epsilon_{4}(\bk) \tau_1
\alpha_3 \gamma_{3},
\label{eq:Kinetic terms}
\end{equation}
is a tight-binding description of the hopping terms
generated by direct hopping between the iron atoms, and virtual
hopping via the out-of-plane As (Se) atoms.
Note the absence of the Nambu matrix $\gamma_{3}$ in the
second and third term of the hopping Hamiltonian: this is because
the orbital and tetrahedral operators are invariant under the particle
hole transformation,
$\vec{\alpha} \rightarrow - \lambda_{2}\vec{\alpha}^{T}\lambda_{2}= \vec{\alpha} $,
$\vec{\tau }\rightarrow - \lambda_{2}\vec{\tau}^{T}\lambda_{2}= \vec{\tau} $.
Here $\epsilon_{1}(\bk) = 4 t_1 (c_{x}+c_{y})$
and $\epsilon_{0}(\bk) = 4 t_0 c_{x}c_{y}-\mu$
are the orbitally-independent amplitudes for
nearest and next nearest-neighbor hopping,
where $\mu$ is the chemical potential and $c_{l}\equiv \cos k_{l}$
($l=x,y$).
The terms $\epsilon_{4}(\bk) = 4 t_4 (c_{x}-c_{y})$
and $\epsilon_{3}(\bk) = 4 t_3 s_{x}s_{y}$
are the amplitudes for
orbitally-dependent nearest and next-nearest neighbor hopping, 
where $s_{l}\equiv \sin k_{l}$ ($l=x,y$).
These terms can be regarded as orbital ``Rashba'' fields, which
split the Fermi surface into separate electron and hole pockets 
{with non-trivial band topology\cite{DHLeePRB2009}}.
The resulting normal state spectrum is given by
$\epsilon_{s \alpha} (\bk ) = \mathcal{E}_s (\bk)
+\alpha \sqrt{\epsilon_{3}
(\bk )^{2}+\epsilon_{4} (\bk )^{2}}$, where
$\mathcal{E}_{s} = \epsilon_0(\bk) + sgn(s) \epsilon_1(\bk)$
and $s, \alpha =\pm 1$ are two sets of band indices.
Since $\epsilon_{3} (\bk )= 4t_{3}d_{xy} (\bk )$ and
$\epsilon_{4} (\bk )= 4 t_{4}d_{x^{2}-y^{2}} (\bk )$ have
d-wave symmetry, the product of the these terms
generates matrix elements that {inevitably} hybridize
the two locally d-wave components of an $A_{1}$ or $A_{2}$ tao pair.

Performing a mean-field decoupling of the interaction, we obtain
\begin{equation}
H = \sum_{\bk }\Psi\dg_{\bk }{\cal H} (\bk )
 \Psi_{\bk } + N_{s}\left(\frac{\Delta_{{1}}^{2}}{g_{1}}+\frac{\Delta_{{2}}^{2}}{g_{2}}
 \right)
\label{eq:BCS Hamiltonian}
\end{equation}
where
\begin{equation}\label{}
{\cal H} (\bk ) = h (\bk )+
\Delta_{xy} (\bk )
\tau_2 \alpha_3 \gamma_{2}
+
\Delta _{x^2-y^2}(\bk) \tau_3 \alpha_1\gamma_{1}
\end{equation}
is the Nambu Hamiltonian.
${\cal H} (\bk )$ can be diagonalized, giving rise to four
separate quasi-particle eigenvalues for the
electron and hole pockets, given by
\begin{eqnarray}
\label{eq:energy eigenvalue}
E^{s \alpha}_{\bk } & = & \bigg[ A_s (\bk ) + \alpha \sqrt{A_s (\bk )^2 - B_s (\bk )^2}
\bigg]^{1/2},
\end{eqnarray}
where $s,\alpha = \pm 1$ and
\begin{eqnarray}\label{eq:constants}
A_s & = & \mathcal{E}_s^2 + \epsilon _3^2 + \epsilon _4^2 + \Delta_{xy}^2 + \Delta_{x^{2}-y^{2}}^2, \\ \nonumber
B_s^2 & = & (A_{s} - 2\epsilon_3 ^2 - 2\epsilon_4^2)^2+
            4 (\Delta_{xy} \epsilon_3 - \Delta_{x^2-y^2} \epsilon_4)^2,
\end{eqnarray}
and for clarity we have suppressed the explicit momentum labels $\bk$.
From (\ref{eq:energy eigenvalue}) and (\ref{eq:constants}) we see that $(E^{s+}_{\bk}
E^{s-}_{\bk})^2 = B_s (\bk)^2 >0$ is positive definite, so that the excitation
spectrum is fully gapped. However, the degree of anisotropy is
strongly dependent on the chirality
$\nu= -{\rm sgn} (\Delta_{1}\Delta_{2})$ of the gap:
for $\nu <0 $, ($\Delta_{1}, \Delta_{2}$  in phase),
the second term in $B_{s}$ vanishes at points on the Fermi surface,
so the gap is highly anisotropic, almost closing
with a tiny minimum value
$\Delta_{min}\sim {\rm min}(\Delta_{1}, \Delta_{2})^{2}/W$, where $W$ is
the bandwidth; for positive chirality $\nu >0 $ ($\Delta_{1}
, \Delta_{2}$ out-of-phase), all terms in
$B_{s}^{2}$ remain positive, and the
gap is weakly anisotropic, with a large minimum value
$\Delta_{min} \sim {\rm min} (\Delta_{1},\Delta_{2})$. The weakly anisotropic
positive chirality state is thus kinetically favored. The winding number of the tao state
is given by Eq.~\ref{eq:topological number}, with $\nu=\pm 2$.

\begin{figure}[bht!]
\begin{center}
\includegraphics[trim=0mm 0mm 0mm 1mm, clip, width=\columnwidth]{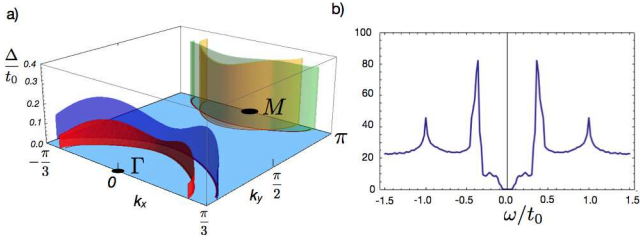}
\end{center}
\caption{(a) SC gap of hole and electron pockets
around $\Gamma$ and $M$.
(b) SC density of states (DOS)
showing clear gap. Energies are measured in units of $t_{0}$.
The parameters used are $t_0 = 1.0, t_1 = 0.1, t_3 = 1.0, t_4 = 0.6, \mu = 0.6, \Delta_1 = 0.5, \Delta_2 = -0.2$.}
\label{Fig:3D SC plot 1}
\end{figure}

\begin{figure}[bht!]
\begin{center}
\includegraphics[trim=0mm 0mm 0mm 1mm, clip, width=\columnwidth]{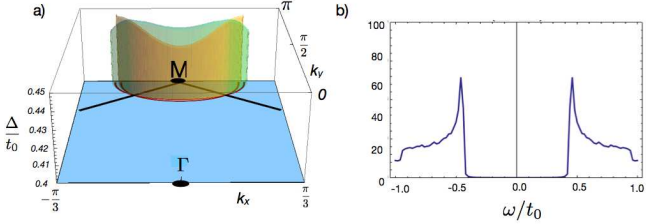}
\end{center}
\caption{(a)SC gap around uncompensated
electron pockets at $M$ points.
Black
lines denote the Brillouin zone. (b)The DOS
for the SC state, showing large gap. Energies are measured in units of
$t_{0}$. Parameters used are $t_0 = 1.0, t_1 = 0.15, t_3 = 0.7, t_4 = 0.7, \mu = -4.0, \Delta_1 = 0.5, \Delta_2 = -0.25$.}
\label{Fig:3D SC plot 2}
\end{figure}

Figs.~\ref{Fig:3D SC plot 1}~\&~\ref{Fig:3D SC plot 2}
show the superconducting gaps for the two generic
cases seen in the iron-based superconductors; two
electron and hole pockets around $M$ and $\Gamma$, and also
two electron pockets around $M$, showing the fully gapped structure in both cases.
A mean field calculation shows that the on-site $s$-wave component induced by the
$A_1$ tao state is less than $1/10$-th of the $d$-wave components
for both the hole and electron pockets and electron pocket only systems;
thus minimizing the Hubbard interaction on the Fe sites compared to an $s$-wave state of similar magnitude.

To explore the internal Josephson coupling between the two condensates,
we minimize the mean-field Free energy
\begin{equation}\label{eq:Free energy}
F= N_{s}\left[\frac{\Delta_{1}^{2}}{g_{1}}+
\frac{\Delta_{2}^{2}}{g_{2}} \right]
- 2 T \sum_{\bk, n, \alpha } \ln \left[2 \cosh  \left(
\frac{E^{s \alpha }_{\bk }}{2T} \right) \right]
\end{equation}
with respect to $\Delta_{1}$ and $\Delta_{2}$, to obtain two coupled gap
equations
\begin{eqnarray}\label{l}
\begin{pmatrix}
1/g_{1}
- \chi_{xy} & - \chi_{J}\cr
- \chi_{J} & 1/g_{2}- \chi_{x^{2}-y^{2}}
\end{pmatrix}
\begin{pmatrix}\Delta_1 \cr \Delta_{2}\end{pmatrix}=0
\end{eqnarray}
where, denoting $\int_{\bk }\equiv
\int
\frac{d^{2}k}{(2\pi)^{2}}$ 
,
\begin{equation}\label{}
\chi_{xy} =
\sum_{s\alpha }\int_{\bk}
\frac{\rm th(
\beta E_{\bk }^{s\alpha }/2)}
{2 E_{\bk }^{s\alpha }}
\left[1+ 2\alpha \frac{(4 t_{4}d_{x^{2}-y^{2}})^{2}}{\sqrt{A_{s} ^{2}-
B_{s} ^{2}}
}
\right]d^{2}_{xy}
\end{equation}
\begin{equation}\label{}
\chi_{x^{2}-y^{2}} =
\sum_{s\alpha }\int_{\bk }
\frac{\rm{th}(
\beta E_{\bk }^{s\alpha }/2
)}
{2 E_{\bk }^{s\alpha }}
\left[1+ 2\alpha   \frac{(4t_{3}d_{xy})^{2}}{\sqrt{A_{s} ^{2}-
B_{s}^{2}}
}
\right]d^{2}_{x^{2}-y^{2}}
\end{equation}
are the
$d_{xy}$ and $d_{x^{2}-y^{2}}$ pair-susceptibilities,
 while
\begin{equation}
\chi_{J} = \sum_{s\alpha }\alpha
\int_{\bk}
\frac{
\rm{th}(
\beta E_{\bk }^{s\alpha }/2
)}
{2 E_{\bk }^{s\alpha }}
\left[
\frac{ 2t_{3} t_{4}
}{\sqrt{A_{s} ^{2}-
B_{s} ^{2}}
} \right] (4d_{xy}d_{x^{2}-y^{2}})^{2}
  \label{eq:Chi Josephson}
\end{equation}
is the internal Josephson coupling between them.
Since  the $E^{s\alpha }_{\bk }$ have the lowest magnitude when
$\alpha = -1$, $\chi_{J}$ is negative, inducing a $\pi$ coupling
between the d-wave condensates, favoring selection of a positive chirality $\nu>0$.
The mean-field transition temperature is determined by
$(1/g_{1}-\chi_{xy}) (1/g_{2}-\chi_{x^{2}-y^{2}})- \chi_{J}^{2}=0$.
In a conventional d-wave superconductor, pairing develops exclusively
in the single most attractive d-wave channel, but in this system, the
strong Josephson coupling
between the two states will immediately lock the two gaps
into a topological state with a single, enhanced transition temperature.

We end with a discussion of the experimental consequences
of tao-pairing.
There are a number of well established features of the experiments,
such as the the absence of a Hebel-Slichter peak in NMR measurements
\cite{EisakiTerasakiJPSJ09,EisakiYashimaJPSJ09}, and resonance peaks in the
neutron scattering \cite{HinkovNatPhys10,GuidiNat08}
that are qualitatively consistent with the $d$-wave character of tao
pairing.
In multi-layer iron systems, tao pairing could occur in four
different global symmetry representations, $A_{1}$, $A_{2}$ and
the staggered $A^*_{1}$ and $B^*_{1}$ representations, driven by
interlayer physics.
The uniform $A_{1}$ ($s$-wave) symmetry
driven by a
strong in-plane Josephson coupling
appears
to be the best candidate: it
allows us to understand
the observed c-axis tunneling into s-wave superconductors \cite{TakeuchiPRL09},
and posseses  the most isotropic
gap,
a feature
consistent with STM experiments \cite{HoffmanRPP11}.

The staggered d-wave  $A_{1}^{*}$ and $B^{*}_{1}$
states may be relevant to systems with strong interplane coupling.
Some experiments suggest a multi-gap character,
including STM \cite{HoffmanRPP11,AllanScience12},
ARPES experiments \cite{DingEPL08}, and the
observation of a cross-over in the NMR relaxation rate from $\tfrac{1}{T_1}$ from $T^3$ at high temperatures
to $T^5$\cite{EisakiTerasakiJPSJ09,EisakiYashimaJPSJ09}
behavior at low temperatures, features
consistent with the $A_{1}^{*}$ or $B_{1}^{*}$ condensates,
where the staggered d-wave gap  $\Delta_{xyz}=
\Delta_{1}s_{x}s_{y}s_{z} $ will lead to out-of-plane point nodes.

%STM experiments also show differing $s$ or $d$-wave like
%DOS in different systems \cite{HoffmanRPP11}. As described above, the phase between
%$\Delta_{xy}$ and $\Delta_{x^2-y^2}$ will cause a large or
%small gap; thereby resulting in an $s$-wave like DOS, or a DOS with a very small gap that appears $d$-wave like.
%Our theory therefore naturally accounts for the multi-gap behavior seen experimentally.

An interesting feature of the theory is the
internal chirality $\nu$ of the pairs.
Pairs of reversed chirality, generated by anti-phase
fluctuations of the two gaps are expected to lead
to a low-lying Raman-active Leggett mode \cite{Leggett66,Blumberg07}
There may also be
higher-energy excitations of Cooper pairs of a different symmetry, for
e.g. from $A_{1}$ to $B_{1}$, which will couple to
corresponding Raman modes.
The emergence of these low-lying ``failed d-wave''
excitations below $T_{c}$ is key property of the tao-condensate.

The chirality of the gap gives it a topological character.
Symmetry  analyses of the pairing Hamiltonian
shows that it lies in the ``CI'' class of topological superconductors
\cite{Schnyder:2008ez}. In two dimensions, these states are
topologically trivial, but their
three dimensional extension
contains a
topological integer ($Z$) invariant which allows for the
possibility of non-trivial surface Andreev states.
%In the absence of the symmetry breaking
%terms $\epsilon_{1,3,4}$ of the kinetic Hamiltonian, the fully
%degenerate paired state will exhibit a gapless Andreev edge mode
%in a fashion analogous to $^{3}$He-B.
%The reinstatement of the additional terms is analogous to introducing
%magnetic fields in the tetrahedral and orbital channels that would give a mass
%to the gapless edge modes. However, it preserves the gap and the topology of the state,
%breaking the degeneracy between $\nu= \pm $ but without breaking
%either lattice or time-reversal symmetry.

%% Open questions.
%Several important and interesting issues remain open.

% Rephrase.
%In conclusion, symmetry analysis of the
%iron-based superconductors lead us to predict the existence of a new
%class of spin singlet, tetrahedral and orbital triplet pairing (``tao pairing''),
%in which a topologically non-trivial combination of locally $d_{xy}$ and $d_{x^{2}-y^{2}}$
%paired states gives rise to a fully gapped excitation spectrum.
%This novel state utilizes two new quantum iso-spins, providing an interesting new
%candidate symmetry for the fully gapped iron-based superconductors.

We gratefully acknowledge discussions with Girsh Blumberg,
J. C. Seamus Davis, Matthew Foster, Gabriel Kotliar,
Joerg Schmalian, and Manfred Sigrist on aspects of this work. This work is supported by DOE grant DE-FG02-99ER45790.

\bibliography{FeAs_Bib}

\end{document}